\documentclass[a4paper,11pt]{article}
\pdfoutput=1
\usepackage{jcappub} 

\usepackage[T1]{fontenc} 

\newcommand{\be}{\begin{equation}}
\newcommand{\ee}{\end{equation}}
\newcommand{\ba}{\begin{eqnarray}}
\newcommand{\ea}{\end{eqnarray}}
\newcommand{\n}[1]{\label{#1}}
\newcommand{\eq}[1]{Eq.(\ref{#1})}

\newcommand{\hh}{\, ,\hspace{0.5cm}}

\newcommand{\BM}[1]{{\mbox{\boldmath $#1$}}}

\title{Spectral line broadening in magnetized black holes}
\author{Valeri P. Frolov,}
\author{Andrey A. Shoom}
\author{and Christos Tzounis}
\affiliation{Theoretical Physics Institute, University of Alberta,\\
Edmonton, AB, Canada,  T6G 2E1}
\emailAdd{vfrolov@ualberta.ca}
\emailAdd{ashoom@ualberta.ca}
\emailAdd{tzounis@ualberta.ca}

\abstract{We consider weakly magnetized non-rotating black holes.
In the presence of a regular magnetic field the motion of charged particles in the vicinity of a black hole is modified. As a result, the position of the innermost stable circular orbit (ISCO) becomes closer to the horizon. When the Lorentz force is repulsive (directed from the black hole) the ISCO radius can reach the gravitational radius. In the process of accretion charged particles (ions) of the accreting matter can be accumulated near their ISCO, while neutral particles fall down to the black hole after they reach $6M$ radius. The sharp spectral line Fe K$\alpha$, emitted by iron ions at such orbits, is broadened  when the emission is registered by a distant observer. In this paper we study this broadening effect and discuss how one can extract information concerning the strength of the magnetic field from the observed spectrum.}


\begin{document}
\maketitle

\flushbottom

\section{Introduction}

There are many indications that a magnetic field plays an important role in astrophysical black holes \cite{Na:05,Fer:97,FrNo:98,Pu:08,Pi:10}. In particular, the magnetic field is required for the energy transfer from accretion disks to jets, as well as the collimation of the jets themselves. Different models, such as the Blandford-Znajec mechanism \cite{Zn:76,BlZn,MEM} and the Penrose mechanism for a magnetic field \cite{KoShKuMe:02,Ko:04}, agree that in order to explain the produced power of about $10^{45}$erg in jets by a supermassive black hole of mass $10^9 M_{\odot}$, a regular magnetic field of the order of $10^4$G is required. Recent observations of the Faraday rotation of radiation from a pulsar in the vicinity of the Milky Way black hole (SgrA*) implies that one should expect the magnetic field of several hundred Gauss in its vicinity \cite{Ea:13}. Models of supermassive black holes with superstrong magnetic field were discussed in \cite{Kard}.

It is known that a regular magnetic field near the black hole can strongly modify the motion of charged particles \cite{AG,FS,FF}. This effect is controlled by a parameter
\be\n{bbb}
b={qBMG\over mc^4}.
\ee
Here $q$ and $m$ are charge and mass of the charged particle, $B$ is the strength of the magnetic field, and $M$ is the mass of the black hole. In the CGS system of units this parameter $b$ is dimensionless. If $b>1$ the charged particle orbits are quite different from the Keplerian ones. For a proton near a stellar black hole of the  mass $M=10 M_{\odot}$ this parameter takes the value $b=1$ for the magnetic field $B\sim 2$ G. For a supermassive black hole of the mass $M\sim 10^9 M_{\odot}$ the corresponding field is $B\sim 2\times 10^{-8}$ G.  One can expect that the parameter $b$ is large for astrophysical black holes\footnote{It should be emphasized that in order to modify the background geometry the magnetic field must be of the order of $10^{19}(M_{\odot}/M)G$. The magnetic fields which we discuss in this paper are much weaker, so that their backreaction is neglected\cite{AG,FS}.}. For simplicity in this paper we consider non-rotating black holes.

The matter in an accretion disk surrounding a black hole is in the form of a plasma. Particles in the disk move along the Keplerian circular orbits, slowly loosing their energy and angular momentum and approaching the inner edge of the disk. It is located at the radius of the innermost stable circular orbit (ISCO), which for neutral matter is at $6M$. After this, neutral particles almost freely fall into the black hole, bringing to it their energy and angular momentum. However, in the presence of the magnetic field an ISCO radius for charged particles is smaller than $6M$ and can be located close to the gravitational radius $2M$. Hence we assume that charged particles (ions) can be accumulated near their ISCO. If Iron ions are revolving around the black hole at such orbits, the broadening of their sharp emission lines depends both on the radius of the orbit and their velocity.
In this paper we discuss how such information, encoded in the observed spectrum,
can used for getting information about the magnetic field in the black hole vicinity.

The methods  using Fe K$\alpha$ lines as probes of the black hole vicinity are well known and widely discussed in the literature
(see, e.g., \cite{Fabi:89,Stella,Laor,matt,LtoN95,ReBe:77,BrChMi,ZaNuPaIn:05,Za:07, KaMi:04,FaIwRe:00,ReNo:03,Zakh:03,Jo:12} and references therein).
As a result of the fluorescence of K$\alpha$ in a relatively cold accretion disc, such a spectral line is excited. It is intrinsically very narrow. However, when such an atom or ion is revolving around the black hole, the frequency of registered photons depends on the position and velocity of the emitter. As a result, the observed spectrum is broadened. These broadening spectra were calculated for non-rotating and rotating black holes  and the method of calculations can be found in the literature \cite{CuBa:73,Cu:75,Zakh:94,CaFaCa:98,FrKlNe:00,FuWu:04,DeAg:09}. Here we are going to perform similar calculations for magnetized black holes for the following reasons: (1) In the standard calculations for non-rotating black holes it is assumed that the emitter moves along a Keplerian orbit and its radius is $\ge 6M$, while for the magnetized black holes one needs to consider closer orbits (up to $2M$); (2) Even for the orbit of the same radius, the angular velocity in magnetized black holes differs from the Keplerian one; (3) In the standard calculations the averaging over the total disk surface is performed, while in our set-up one can expect that the region where the charged particle motion is modified by the magnetic field  is much narrower that the disk size.

This paper contains a brief summary of the results. The paper containing all the details of the calculations will be published elsewhere \cite{FST}. In this paper we use units where $c=G=1$.

\section{Charged particles in magnetized black holes}

\subsection{Equations of motion}

In this paper we use the following model. We consider a non-rotating black hole of the mass $M$ and write its metric in the form
\be\n{met}
dS^2=r_g^2 ds^2\hh
ds^2=-f dt^2+{d\rho^2\over f}+\rho^2 (d\theta^2+\sin^2\theta d\phi^2)\, .
\ee
Here $r_g=2M$ is the gravitational radius,  $f=1-1/\rho$ and $t$ and $\rho$ are dimensionless time and radial coordinates, respectively. We write the potential of the magnetic field in the form (see, e.g., \cite{Wald})
\be
\BM{A}={B r_g\over 2}\partial_{\phi}\, .
\ee
Such a field is axisymmetric and homogeneous at infinity where its strength is $B$.

\begin{figure}[htb]
\hfill
 \includegraphics[width=6cm]{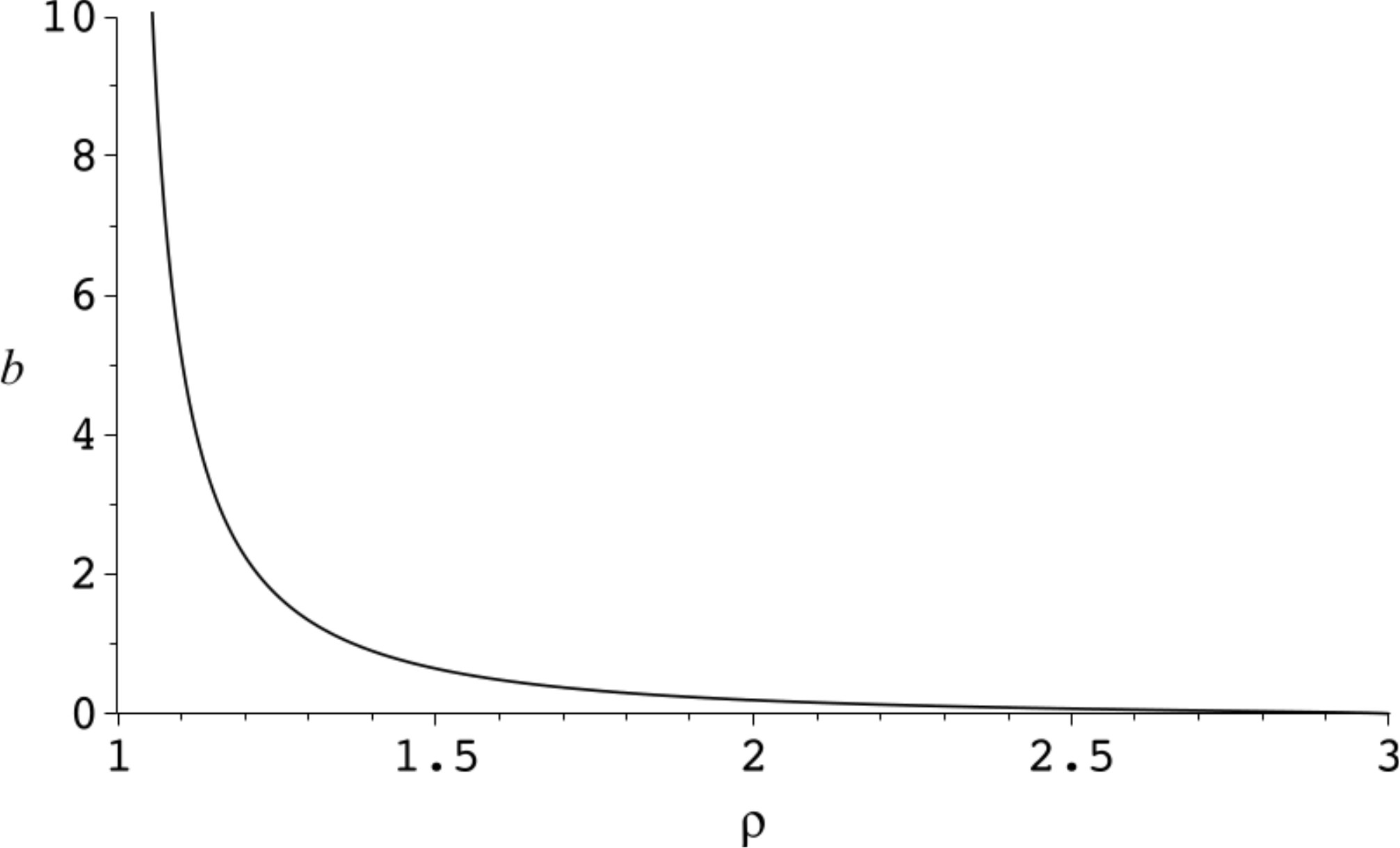}
\hfill
  \includegraphics[width=6cm]{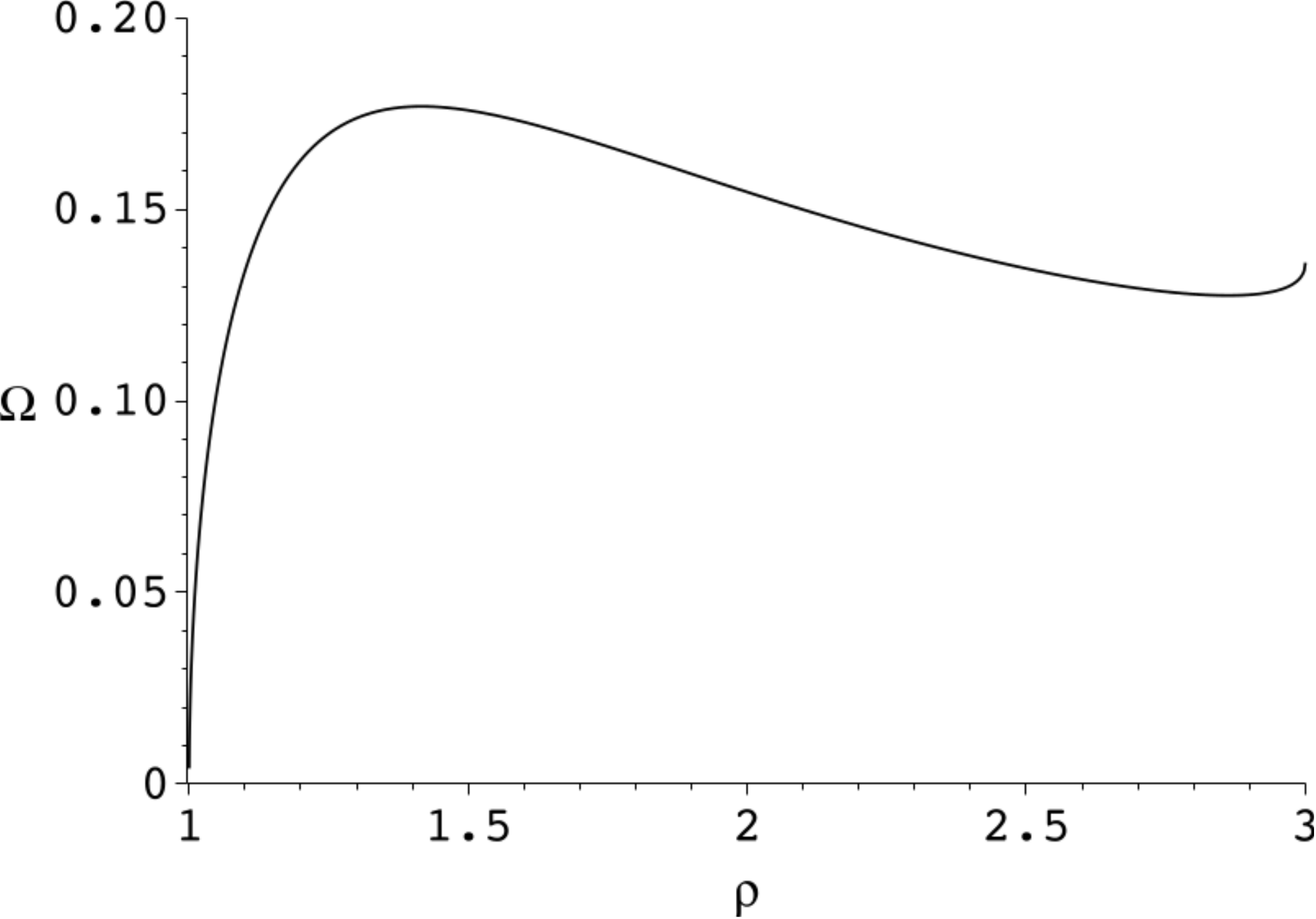}
  \hfill  \, { }
  \caption{ \label{Fig_2}   The left plot shows the relation between the radius $\rho$ of the ISCO of the charged particle and the value of the magnetic field $b$. The right plot shows how the angular velocity $\Omega$ of the particle at ISCO depends on its radius. \label{Fig_1}}
\end{figure}

\begin{figure}[htb]
\begin{center}
  \includegraphics[width=7cm]{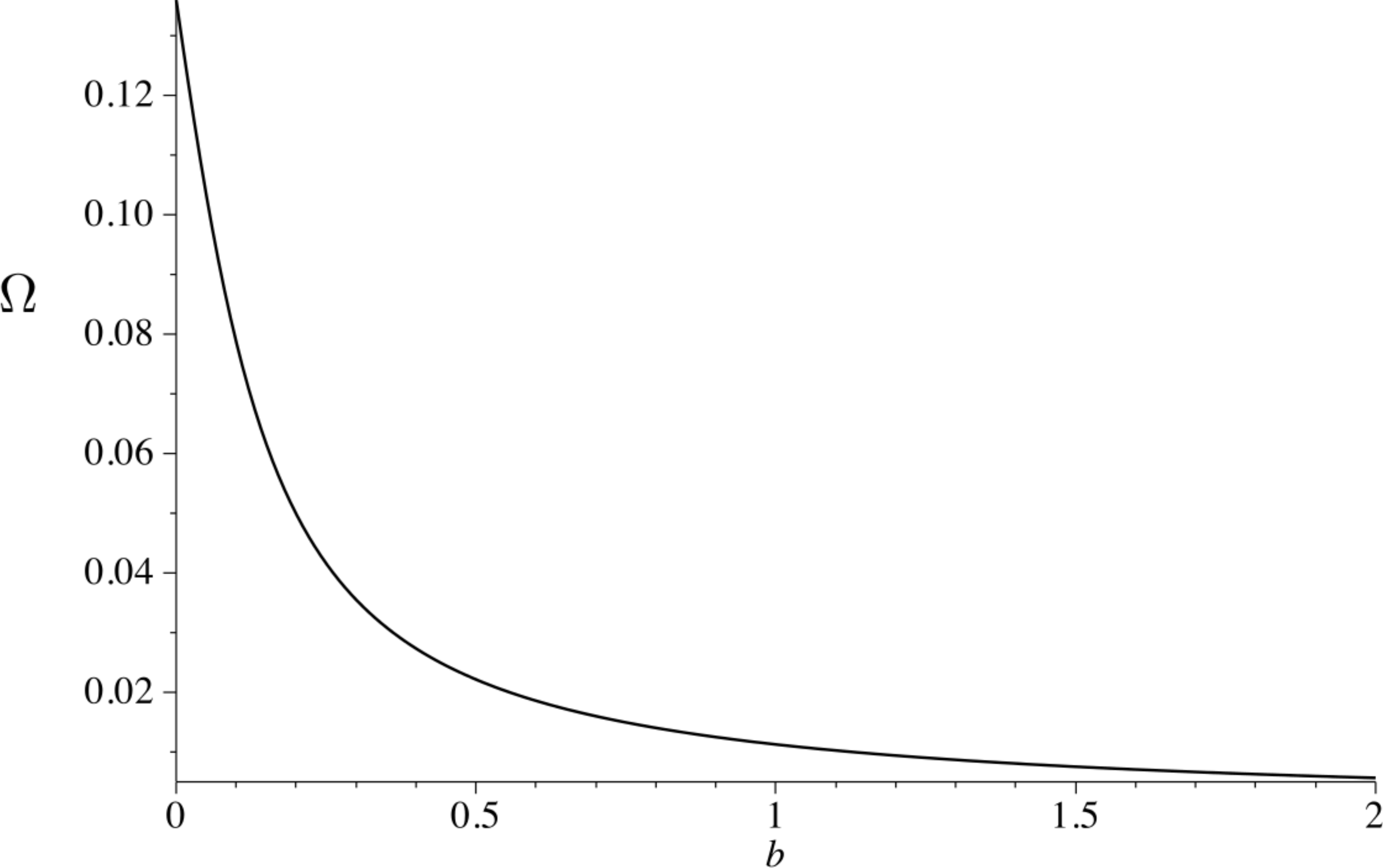}\\
  \caption{This figure shows how the angular velocity $\Omega$ of the particle at the fixed circular orbit at $\rho=3$ depends on the value of the magnetic field $b$. \label{ANG_VEL}}
  \end{center}
\end{figure}

We assume that a charged particle (mass $m$, charge $q$) moves in the equatorial plane, that is, in the plane $\theta=\pi/2$ orthogonal to the magnetic field. Its equations of motion are completely integrable and can be written in the first order form
\ba
&&\hspace{0.8cm}\left(\frac{d\rho}{d\tau}\right)^2={\cal E}^2-U\,,\n{12}\\
&&\frac{d\phi}{d\tau}=\frac{{\cal L}}{\rho^2}-b
\hh {dt\over d\tau}=\frac{{\cal E}\rho}{\rho-1}\,.\n{13}
\ea
Here ${\cal E}$ is the conserved specific energy (energy per a unit mass) and ${\cal L}$ is conserved  generalized specific azimuthal angular momentum. We focus on the case when ${\cal L}>0$ and the Lorentz force is repulsive, that is, it is directed from the black hole. Namely this case has the most interesting applications\footnote{For a discussion of a general case see, e.g., \cite{FS}.}. The parameter $b$  is a dimensionless strength of the magnetic field at infinity and it is determined by \eq{bbb}.

The effective potential
\be\n{14}
U=\left(1-\frac{1}{\rho}\right)
\left[1+\frac{({\cal L}-b\rho^2)^2}{\rho^2}\right]\,.
\ee
determines main features of the radial motion. It has the following properties: $U$ grows as $\rho^2$ at infinity. This means that a charged particle with finite energy cannot reach infinity. The equation ${\cal E}=\sqrt{U}$ determines the maximal radius where such a particle has a radial turning point. In other words, in the presence of the magnetic field all orbits of charged particles are bounded. The potential $U$ vanishes at the horizon $\rho=1$. It either has one local maximum and one local minimum, or it is a monotonic function. We consider stable circular orbits which are determined by a condition $dU/d\rho=0$. For a given magnetic field $b$ such orbits exist when $\rho\ge \rho_{ISCO}$.

\subsection{Innermost stable circular orbits}

The radius $\rho_{ISCO}$ of the innermost stable circular orbit is determined by the conditions
\be\n{eeqq}
{dU\over d\rho}={d^2U\over d\rho^2}=0\, .
\ee
These two equations establish relations between three quantities $\rho_{ISCO}$, $b$, and ${\cal L}$. One can solve \eq{eeqq} analytically and find the explicit expressions for $b$ and ${\cal L}$ as functions of $\rho_{ISCO}$. One also has ${\cal E}=\sqrt{U(\rho_{ISCO})}$.  Since the corresponding expressions are rather long, we do not present them here. They can be found in \cite{FS,FST}. Here we discuss some important properties of ISCO's in magnetized black holes. In the absence of the magnetic field the radius of ISCO is $\rho= 3$ (in $r_g$ units) and the specific energy is ${\cal E}_{ISCO}=\sqrt{8}/3$. In the presence of the magnetic field the ISCO for charged particles is closer to the horizon. Its radius is monotonically decreasing function of $b$ and, in the limit $b\to \infty$, it tends to the horizon ($\rho=1$) (see the left plot in Figure~\ref{Fig_1}). The specific energy ${\cal E}_{ISCO}$ monotonically decreases with the increase of $b$ and in the limit $b\to\infty$ when $\rho_{ISCO}\to 1$ one has ${\cal E}_{ISCO}\to 0$. This property means that the efficiency of the energy extraction is much higher for charged particles in magnetized black holes (up to $100\%$) than for neutral ones (up to $5.7\%$). The right plot in Figure~\ref{Fig_1} shows the dependence of the angular velocity of the ISCO particle on the radius. One can see, that in the presence of the magnetic field the angular velocity of ISCO particles  becomes smaller and vanishes in the limit $b\to\infty$. This is a result of a very strong redshift effect in the vicinity of the horizon.

The action of the magnetic field on the angular velocity can be illustrated in a slightly different way. In the absence of the magnetic field the Keplerian angular velocity at ISCO is $1/(3\sqrt{6})$. Consider now a charged particle in the magnetic field which has the same radius of its circular orbit. Let us stress, that this orbit is not ISCO any more. The plot presented in Figure~\ref{ANG_VEL} shows how the angular velocity $\Omega$ of such a particle depends on the magnetic field. Namely, with the increase of $b$ the angular velocity decreases and tends to zero value. The explanation of this phenomenon is quite simple. The repulsive Lorentz force reduces the required centrifugal acceleration of the revolving particle.

Neutral particles of an accretion disk located in the equatorial plane fall to the black hole practically freely after they reach its inner edge at $6M$. However, charged particles (ions) with positive ${\cal L}$, after passing the radius $6M$, continue to revolve around the black hole slowly losing their energy and angular momentum until they reach their own ISCO radius $\rho(b)$ determined by the parameter $b$. We assume that such particles are accumulated in the domain between $\rho=3$ and $\rho_{ISCO}(b)$. Our purpose is to calculate the spectral broadening of the sharp spectral lines emitted by ions revolving in this domain in the magnetized black hole. Both the position of ISCO and the angular velocity of the charged particles depend on the magnetic field. For this reason one, in principle, can use the dependence of the spectral broadening on the magnetic field parameter $b$ in order to estimate the magnetic field itself.

\subsection{Model}

For calculations of the spectral broadening we use the following simple model. We consider a single ion revolving around the black hole as an emitter sending stationary monochromatic and isotropic (in its own reference frame) radiation. We denote by $\omega_e$ the emitted frequency. We denote by $L \Delta\tau_e$ the total number of quanta emitted  in all directions during the interval $\Delta\tau_e$ of the proper time. We assume that the emitter is revolving in the equatorial plane and its radius and angular velocity are $\rho_e$ and $\Omega_e$, respectively. The azimuthal angle characterizing the position of the emitter is $\varphi=\Omega t_e$. It changes in the interval from $-\pi$ to $\pi$.

Some of the emitted quanta (photons) arrive to the observer located at the far distance $\rho_o$ from the black hole and they are registered by him/her. We define the spherical coordinates $(\theta,\phi)$ so that  $\phi_o=0$ stands for the distant observer. We denote by $\theta_o$ the inclination angle, that is the angle between the direction to the observer and the magnetic field at the infinity. A photon emitted at the time $t_e$ will arrive at some moment $u_o(t_e)$ of the retarded time and it will have frequency $\omega_o(t_e)$. Both of these quantities can be found by solving the photon's equation of motion between the emitter and the observer. In particular, the frequency $\omega_o$ depends on the position and the velocity of the emitter, and hence, contains the information about both  Doppler shift and gravitational redshift effects. For a given $L$ one can find how many photons will be registered by a distant observer in the time interval $(u_o,u_o+\Delta u_o)$ by a device with an effective aperture $A$. If one does not fix the time of arrival of photons, but registers only their frequency, one can determine the quantity $dN_o/d\omega_o$ as a function of the frequency $\omega_o$. We call this quantity the spectrum of the radiation from the emitter\footnote{One can arrive to the same function by assuming that instead of a single ion, there exist many of such ions at the circular orbit of the same radius $\rho_e$. In such a case, an averaging over the angle $\varphi_e$ is effectively equivalent to the integrating (averaging) over the arrival time $u_o$. So that, one again arrives to the same spectral function $dN_o/d\omega_o$.}.

\section{Photon trajectories}

To relate parameters of the emitter and properties of the observed quanta, let us consider propagation of the emitted photons in the Schwarzschild spacetime. The presence of the magnetic field evidently does not affect their motion. It should be emphasized that a similar problem of the spectral line broadening was studied earlier, both in the Schwarzschild and Kerr metrics (see, e.g., discussion in \cite{CuBa:73,Cu:75,FuWu:04,DeAg:09}). However,  since  the ISCO radius $\rho_e$ and angular velocity $\Omega_e$ depend on the magnetic field, for magnetized black holes one needs to perform these calculations again.

We denote by $\Phi$ the angle between the direction to the emitter and the direction to the distant observer. Simple calculations give
\be
\cos\Phi=\cos\varphi \sin\theta_o\, .
\ee
The angle $\varphi$ is a position of the emitter on its circular orbit. We choose $\varphi=0$ for the direction to the observer and assume that $\varphi\in (-\pi,\pi)$.
It is easy to see that the angle $\Phi$ changes in the interval $(\pi/2-\theta_o,\pi/2+\theta_o)$. It takes its minimal value for $\varphi=0$, and the maximal value when $\varphi=\pm\pi$. At $\varphi=\pm\pi/2$ one has $\Phi=\pi/2$.

A motion of a photon in a spherically symmetric spacetime is planar, that is its trajectory lies in a plane $\Pi$ passing through the emitter $P_e$ and the direction to the observer $P_o$. For a given point of emission, the photon trajectory is uniquely determined by only one parameter $\ell$, which is the ratio of the total angular momentum of the photon to its conserved energy (as measured at infinity). A photon moving in the black hole geometry \eq{met} can have no more than one  radial turning point. Depending on the radius of the emitter's orbit and the position $\varphi$ on this orbit, there can be two classes of the null rays connecting $P_e$ and $P_o$. We call them direct (without a radial turning point on their way from $P_e$ to $P_o$), and indirect (which have a radial point along this way). In the further discussion of the properties of the null rays, the formulas become much simpler if instead of the radius $\rho$ one uses the inverse radius $\zeta=\rho^{-1}$. For a direct ray the bending angle between the point of the emission $\zeta_e$ and the point of the observation at infinity $(\zeta=0)$ is
\be\n{Bp}
B_+(\ell;\zeta_e)=\int_0^{\zeta_e} {d\zeta \over \sqrt{\ell^{-2}-(1-\zeta)\zeta^2}}\, .
\ee
An indirect ray moves at first to the smaller value of $\rho$. Only after it passes through the minimal radius $\rho=\rho_m$, it propagates to the observer $P_o$. For such a ray the bending angle can be written in the form
\be\n{Bm}
B_-(\ell;\zeta_e)=2C(\zeta_m)-B_+(\ell;\zeta_e)\, ,
\ee
where
\be\n{CC}
C(z)=2\int_0^{1}{dy\over \sqrt{Z}}\hh
Z=2-y^2-3z+3z y^2-z y^4\, .
\ee
The following equation
\be\n{BBF}
B_{\pm}(\ell,\zeta_e)=\Phi\equiv\arccos(\cos\varphi\sin\theta_0)\, ,
\ee
establishes a relation between the position (angle $\varphi$) of the emitter on the orbit $\zeta_e$ and the angular momentum $\ell$ of the photon that reaches a distant observer with the inclination angle $\theta_o$. In this relation one needs to choose the sign $+$ for a direct ray and $-$ for an indirect one.

An analysis shows that for a given inclination angle $\theta_o$ there exists a limiting value $\rho_*$ of the orbit's radius, such that for $1<\rho_e<\rho_*$ all the rays that reach the distant observer are direct ones. The  critical inverse radius $\zeta_*$ can be found from the equation
\be
C(\zeta_*)=\pi/2+\theta_o\, .
\ee
A solution of this equation monotonically increases from 0 at $\theta_o=0$ to $\zeta_{*,max}\approx 0.568$ at $\theta_o=\pi/2$. When $\zeta_e<\zeta_*$ a circular orbit contains two parts. For the first part, determined by the relations
\be
-\varphi_*<\varphi<\varphi_*\hh \varphi_*=\arccos\left({\cos(C(\zeta_e))\over  \sin\theta_o}\right)\, ,
\ee
one has direct rays, while for the other part of the orbit the rays are indirect.

\section{Spectral line broadening}

\subsection{Redshift}

Equation (\ref{BBF}) determines the angular momentum $\ell$ of a photon, that being emitted at $\varphi$ reaches the distant observer. Since the null ray equations are completely integrable, one can use this integral of motion and to determine the value of the momentum of the photon at any point of its trajectory.  In order to do this we first introduce a convenient orthonormal tetrad. The first unit vector of it is directed along the timelike Killing vector $\BM{\xi}_t$,
\be
\BM{e}_t=|\BM{\xi}_t^2|^{-1/2} \BM{\xi}_t=f^{-1/2}\partial_t\, .
\ee
The second unit vector $\BM{e}_{\rho}$ is directed along the radius. The last two unit vectors are $\BM{e}_{\Phi}$ and $\hat{\BM{e}}$. They are orthogonal to the previous two vectors and are tangent to $\rho=$\, const 2-sphere. The first of these vectors, $\BM{e}_{\Phi}$, is directed along the Killing vector $\BM{\xi}_{\phi}$, which is a generator of the rotations preserving the position of the photon's plane $\Pi$. We choose its direction to be from $P_o$ to $P_e$. The last vector of the tetrad, $\hat{\BM{e}}$, is uniquely defined by the property that the tetrad is right-hand oriented.

Denote by $\BM{p}$ the four-momentum of a photon. Then using the property $\BM{p}^2=0$ one finds
\be\n{ppp}
\BM{p}=\omega_{o} \left({1\over \sqrt{f}}\BM{e}_{t}+
{1\over \sqrt{f}}{\cal P}\BM{e}_{\rho}+\ell \zeta \BM{e}_{\Phi}\right)\, ,\
{\cal P}=\sqrt{1-\ell^2 \zeta^2 f}\, .
\ee
Here $\omega_o$ is the photon frequency at infinity, $\omega_o=-(\BM{p},\BM{e}_t)$, and $\ell=\omega_o^{-1}(\BM{p},\BM{e}_{\Phi})$.

\begin{figure}[tbp]
\hfill  \includegraphics[width=5cm]{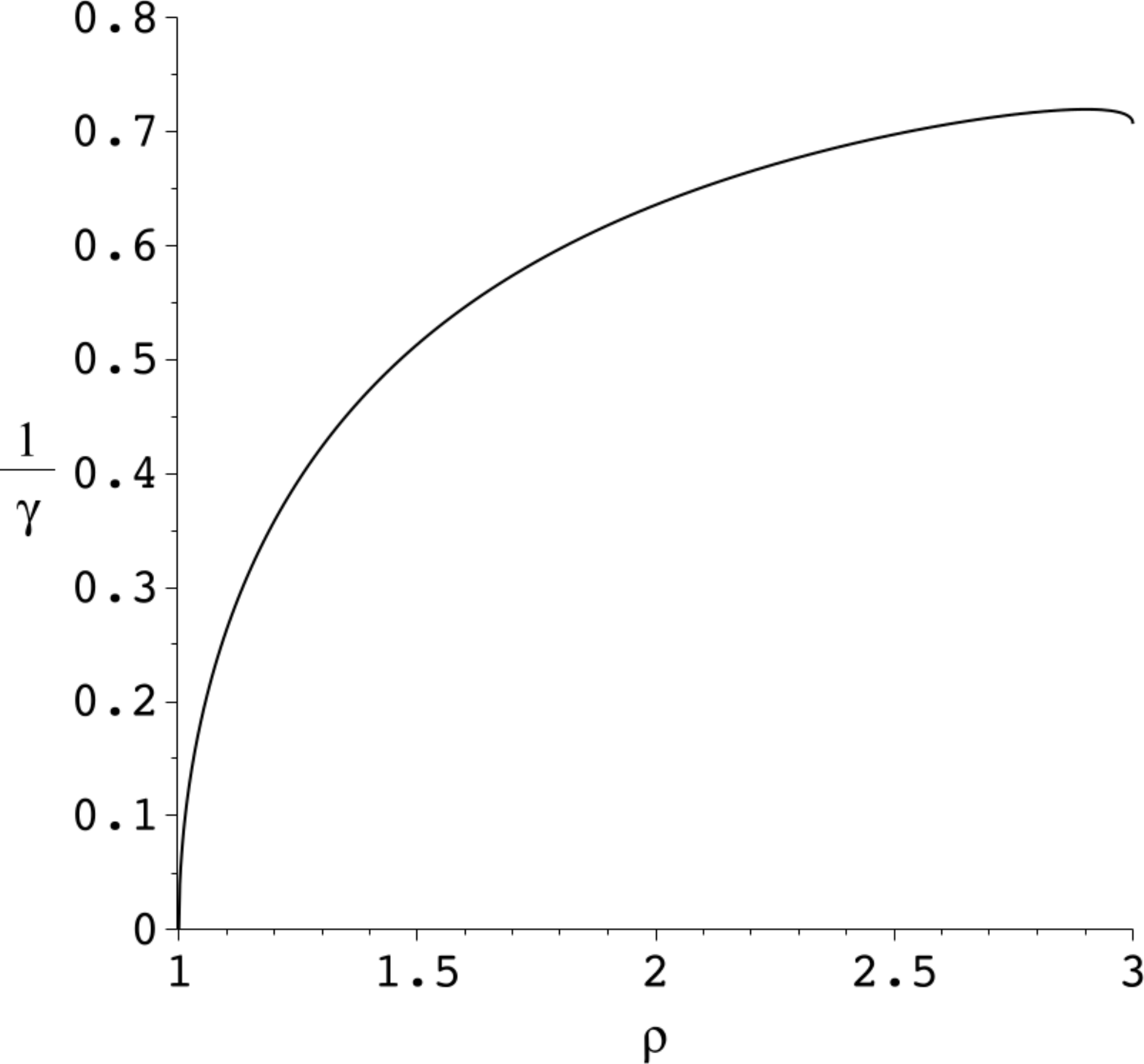}
 \hfill
  \includegraphics[width=5cm]{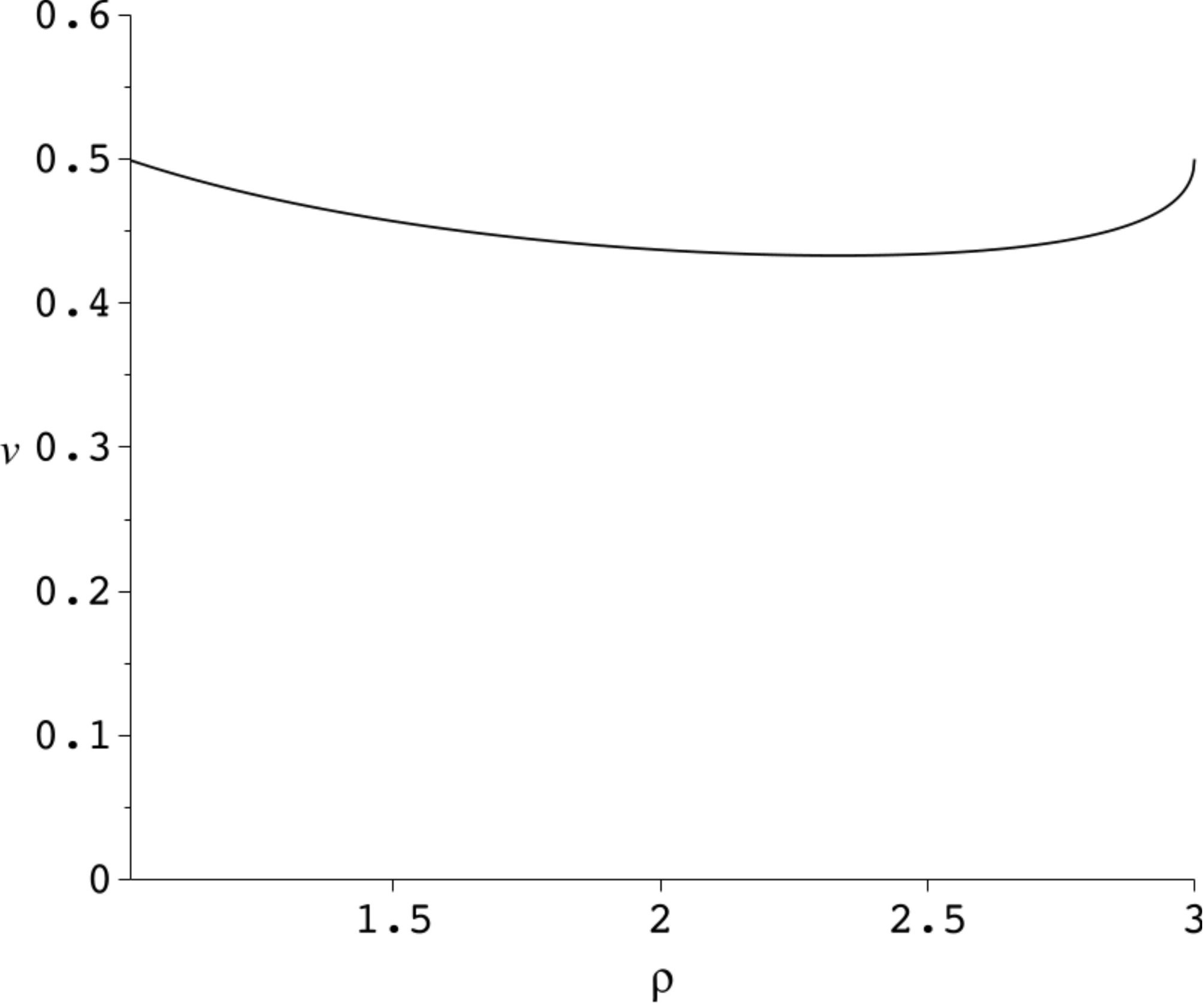}
  \hfill \, { }
  \caption{The left figure shows how the parameter $\gamma$, which controls the redshift effect for the charged particle at the ISCO orbits depends on the value of the magnetic field $b$. The right figure shows how the local velocity $v$ of the charged particle at the ISCO orbits depends on its radius $\rho$. \label{gamma}}
\end{figure}

Let us at first establish the relation between the emitted ($\omega_e$) and registered ($\omega_o$) frequencies. The four-velocity vector of the emitter is
\be
\BM{u}=\hat{\gamma} \left(\BM{e}_t+v\BM{e}_{\phi}\right)\, ,\
v={\Omega\rho\over \sqrt{f}}\, ,\  \hat{\gamma}={1\over \sqrt{1-v^2}}\, ,
\ee
and $\BM{e}_{\phi}$ is a spatial unit vector, tangent to the orbit
\be
\BM{e}_{\phi}=-{1\over \sin{\Phi}} (\sin\theta_o\sin\varphi \, \BM{e}_{\Phi}+\cos\theta_o \, \hat{\BM{e}})\, .
\ee
Using the definition of the emitted frequency $\omega_e=-(\BM{u},\BM{p})$, one obtains
\be\n{alpha}
\alpha={\omega_e\over \omega_o}=\gamma\left(1+{\ell\Omega \sin\varphi\sin\theta_o\over \sin\Phi}\right)\hh
\gamma={\zeta_e\over \sqrt{(1-\zeta_e)\zeta_e^2-\Omega^2}}\, .
\ee

\subsection{Calculation of spectral broadening}

In order to relate the number of emitted photons $L \Delta \tau_e$ and the number of registered ones we proceed as follows. Consider a beam of photons emitted at $P_e$ with momenta slightly different from $\BM{p}$ given by \eq{ppp}. Namely, we consider two independent variations $\delta_{\ell}\BM{p}$ and $\delta_{\psi}\BM{p}$. The first variation preserves the position of the photon's plane $\Pi$, but modifies the photon's trajectory by changing its angular momentum from $\ell$ to $\ell+\delta\ell$. The second variation changes the position of the photon's plane. Namely, it rotates the plane $\Pi$ around the direction to $P_e$ by the angle $\delta\psi$. The corresponding solid angle for such a beam (as measured in the frame comoving with the emitter) can be found by means of the following relation (for the proof see, e.g., \cite{FST}):
\be
\BM{\cal A}=\pm \Delta\Omega_e \BM{E}\, ,
\ee
where $\BM{E}$ is a unit rank-4 totally skew-symmetric tensor and
\be
\BM{\cal A}=\omega_e^{-3} \BM{u}\wedge\BM{p}\wedge\delta_{\ell}\BM{p}\wedge\delta_{\psi}\BM{p}\, .
\ee
Using this relations one finds
\be\n{SSS}
\Delta\Omega_e={\ell \zeta_e^{2} \over \alpha^2 {\cal P}}\delta\ell\  \delta\psi\, .
\ee

If one puts a screen orthogonal to the beam at the distance $\rho_o$ from the black hole, the area of the image of the beam is
\be
A= \rho^2_o B'\sin\Phi\,\delta\ell\, \delta\psi\, ,
\ee
where $B'=\partial_{\ell}B_{\pm}(\ell,\zeta_e)$. One can identify $A$ with the aperture of a `telescope', which registers photons. The number of photons registered per the observer's unit time is
\be
{dN_o\over dt_o} ={d\tau_e\over dt_o}{L\Delta\Omega_e\over 4\pi}\, .
\ee
Using these relations one obtains the spectrum of the observed photons
\be\n{omom}
{dN_o\over d\omega_o}={\Omega \over 2\pi}(dN_o/dt_o)/(d\omega_o/dt_o)\, .
\ee
Here we introduce an additional factor $\Omega/2\pi$ which requires an explanation. The observed frequency $\omega_o$ is a periodic function of $t_o$ with the period $T_o=2\pi/\Omega$. This is a time of the complete revolution of the emitter as measured at infinity. As we shall see later, the frequency $\omega_o$ changes in some interval $(\omega_{min},\omega_{max})$ and  in this interval there exist two branches of the function $\omega_{o}(t_o)$: in the first branch $d\omega_o/dt_o>0$, while in the second one $d\omega_o/dt_o<0$. Denote by $\hat{N}$ the following quantity
\be\n{NPh}
\hat{N}_o=\oint_{\omega_o}{dN_o\over d\omega_o}d\omega_o\, ,
\ee
where the integral is taken over both the branches. This gives the total number of photons received  by the observer during one period of revolution of the emitter divided by the period $T_o$. In order to provide this useful normalization we included the factor $\Omega/2\pi$ in \eq{omom}.

Calculations give
\be\n{SPEC}
{dN_o\over d\omega_o}={\cal C} { \Omega\over 2\pi}\left| {d\alpha\over d\tau_e}\right|^{-1} {\ell \zeta_e^2\over B' {\cal P}\sin\Phi}\hh
{\cal C}={L A\over 4\pi \rho_o^2 \omega_e }\, ,
\ee
where
\be\n{AAA}
{d\alpha\over d\tau_e}={\Omega^2\sin\theta_o \zeta_e^2\over ((1-\zeta_e)\zeta_e^2-\Omega^2)}
\left( {\ell \cos\varphi \cos^2\theta_o\over \sin^3\Phi}+
{\sin^2\varphi \sin\theta_o\over B' \sin^{2}\Phi}\right)\, .
\ee

Let us remind that we use the dimensionless quantities obtained by the rescaling which involves the gravitational radius $r_g$ of the black hole. However, the quantity ${\cal C}$ is scale invariant, so that one can rewrite it in the form
\be\n{CCC}
{\cal C}={\hat{L} \hat{A}\over 4\pi R_o^2\hat{\omega}_e}\, ,
\ee
where the hat means that one restores the correct physical dimensionality. The combination
\be\n{jjj}
{\hat{L} \hat{A}\over 4\pi R_o^2}
\ee
which enters \eq{CCC} has a simple meaning. Consider a flat spacetime and an emitter at rest. Then \eq{jjj} gives the number of particles registered per a unit time by the observer located at the distance $R_o$, provided the aperture of his/her `telescope' is $\hat{A}$.

In our set-up ${\cal C}$ may be considered as a dimensionless normalization constant which does not depend on the position and velocity of the emitter. For convenience, one can put it equal to 1. In such a case we say that we are using the {\em Newtonian normalization}. We also denote
\be
{\cal N}_o={\cal C}^{-1}\omega_{e}^{-1}\hat{N}_{o}\, .
\ee

\section{Results}

\begin{figure}[htb]
\begin{center}
\includegraphics[width=6cm]{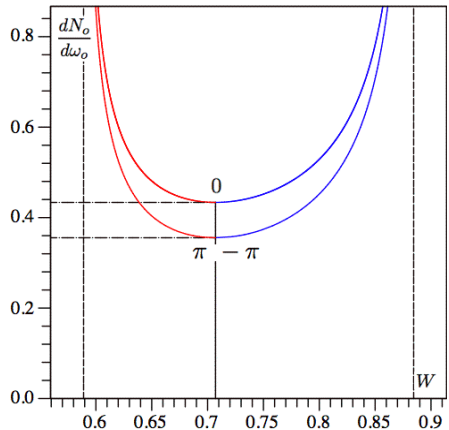}
\caption{Spectral function for ISCO, $b=0$, at $\zeta_{e}=1/3$, 
where the emitter has angular velocity $\Omega=0.136$ and its specific energy is ${\cal E}=0.943$. The total angular momenta $\ell$ of photons that reach the distant observer are in the interval  $[2.898,3.671]$. The spectrum has peaks at $W_{-}=0.589$ ($\varphi_{-}=96^{o}25'$) and at $W_{+}= 0.885$  ($\varphi_{+}=-96^{o}25'$). 
The minimal values ($0.355$ and $0.434$) of $dN_{o}/d\omega_{o}$ for two spectral branches are at $W_{0}=0.707$. One also has ${\cal N}_o=0.371$.}\label{F2}
\end{center}
\end{figure}

\begin{figure}[htb]
\begin{center}
\includegraphics[width=6cm]{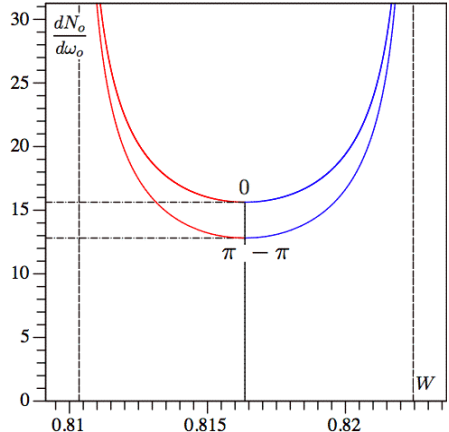}
\caption{Spectral function for SCO, $b=2.251$, at $\zeta_{e}=1/3$. 
The angular velocity of the emitter is $\Omega=0.005$ and its specific energy is ${\cal E}=0.817$. The total angular momenta $\ell$ of photons that reach the distant observer are in the interval  $[2.898,3.671]$. The spectrum has peaks at $W_{-}=0.810$ ($\varphi_{-}=96^{o}25'$) and at $W_{+}= 0.823$  ($\varphi_{+}=-96^{o}25'$).
The minimal values ($12.81$ and $15.63$) of $dN_{o}/d\omega_{o}$ for two spectral branches are at $W_{0}=0.816$. One also has ${\cal N}_o=0.535$.}\label{F3}
\end{center}
\end{figure}

\begin{figure}[htb]
\begin{center}
\includegraphics[width=6cm]{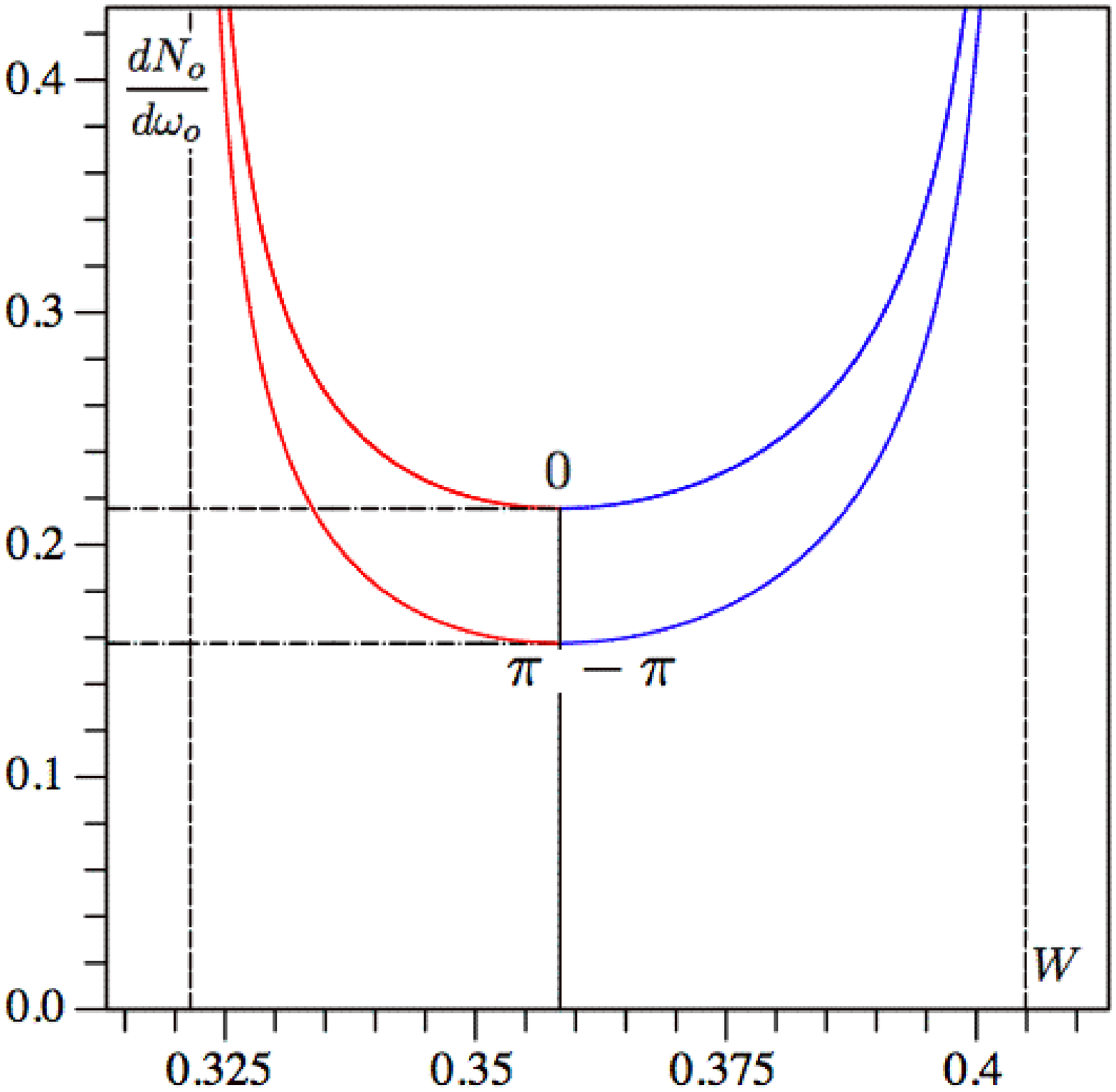}
\caption{Spectral function for ISCO, $b=2.251$, at $\zeta_{e}=5/6$. 
The angular velocity of the emitter is $\Omega=0.163$ and its specific energy is ${\cal E}=0.465$. The total angular momenta $\ell$ of photons that reach the distant observer are in the interval  $[1.260,1.924]$. The spectrum has peaks at $W_{-}=0.322$ ($\varphi_{-}=102^{o}$) and at $W_{+}= 0.405$  ($\varphi_{+}=-102^{o}$).
The minimal values ($0.158$ and $0.216$) of $dN_{o}/d\omega_{o}$ for two spectral branches are at $W_{0}=0.358$. One also has ${\cal N}_o=0.049$.}\label{F4}
\end{center}
\end{figure}

For the calculations of the broadened spectrum one uses the relations (\ref{SPEC}) and (\ref{AAA}). We denote by $W=\omega_o/\omega_e=\alpha^{-1}$ the dimensionless frequency or the redshift factor. Figures~\ref{F2}-\ref{F4} give examples of plots for the spectra. They show $dN_o/d\omega_o$ as a function of $W$. We use there the Newtonian normalization and put ${\cal C}=1$.

\begin{figure}[htb]
\begin{center}
\includegraphics[width=4cm]{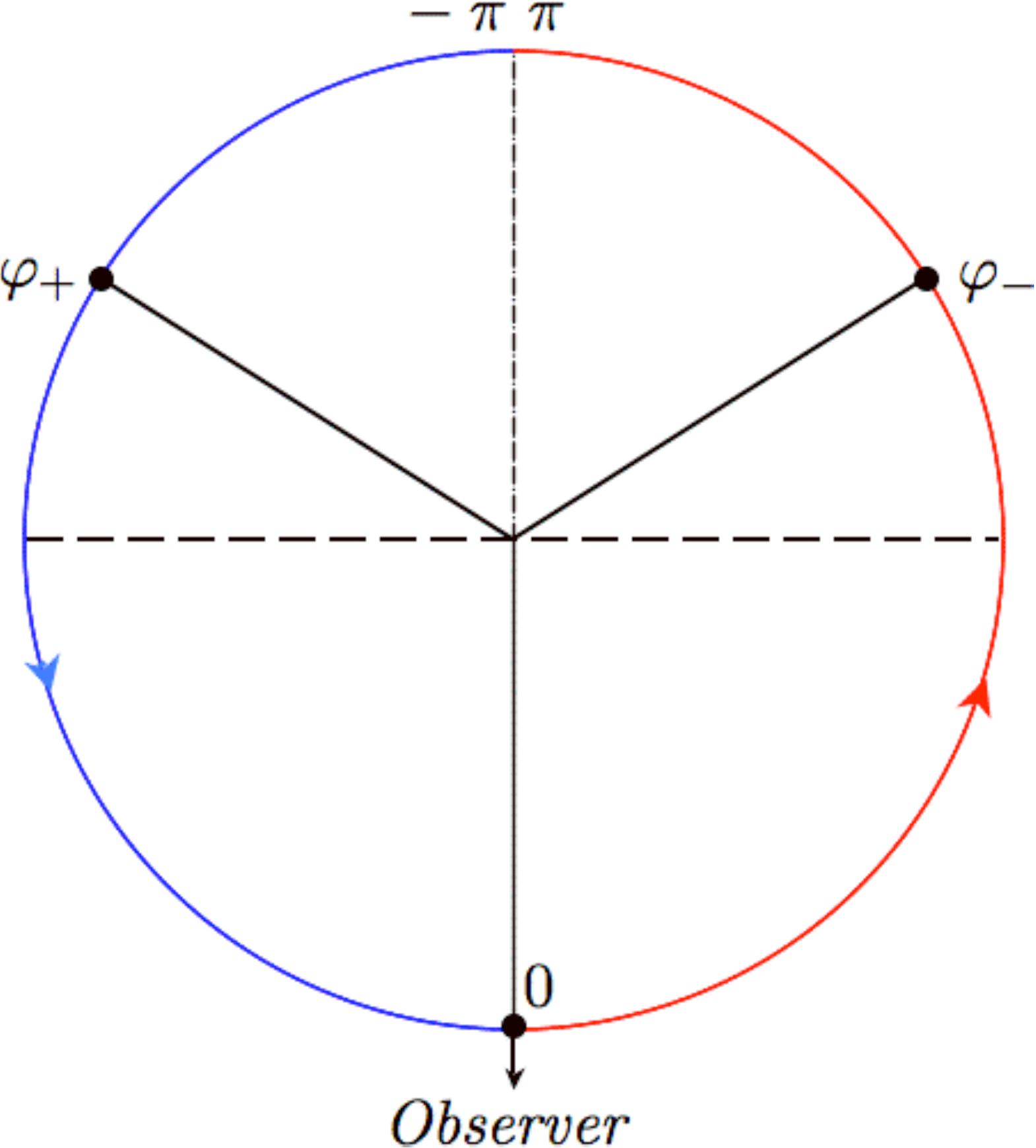}\\
\caption{This diagram schematically shows the orbit of the emitter. The angle $\varphi=0$ corresponds to the direction to the observer. The arrows show the direction of the emitter's motion. For the emitter located in the right semicircle, $\varphi\in[0,\pi]$, photons have Doppler redshift and for the emitter located in the left semicircle, $\varphi\in[-\pi,0]$, photons have Doppler blue-shift. The observed frequency of photons is maximal when they are emitted at $\varphi_+$ and minimal when they are emitted at $\varphi_-=-\varphi_+$. One has $|\varphi_{\pm}|>\pi/2$.} \label{Orbit}
\end{center}
\end{figure}

Before discussing details of these plots let us make the following general remarks. The equation (\ref{BBF}) establishes a relation between the position of the emitter $\varphi$  and the photon's angular momentum $\ell$. It is convenient to solve this equation and express $\varphi$ as a function of $\ell$. For a given orbit, $\ell$ changes in some interval.
Let us consider the expression (\ref{alpha}). Using \eq{BBF} one can rewrite it in the form
\be
\alpha=\gamma [1+\Omega Z(\ell)]\hh
Z(\ell)=\pm{\ell \sqrt{\sin^2\theta_o-\cos^2 B}\over \sin B}\, .
\ee
At the point $\ell_m$, where $|Z(\ell)|$ reaches its maximum value $Z_m$, one has
\be
\alpha=\gamma (1\pm\Omega Z_m)\, .
\ee
At these points $d\alpha/d\ell=0$. We denote
\be
W_{\pm}=\gamma^{-1} {1\over 1\mp\Omega Z_m}\, .
\ee
$W_+$ is the maximal observed frequency of photons. Such photons come  from the emitter when it is at $\varphi_+$. Similarly, $W_-$ is the minimal observed frequency and the corresponding photons are emitted at $\varphi_-$ (see Figure~\ref{Orbit}). At these frequencies the spectral function has peaks. The position of the emitter $\varphi=\Omega t_e$ is a regular (linear) function of time everywhere, including the points where the frequency $W$ reaches its extrema and hence $dW/dt_e=0$ at these points. When one transforms the rate of emission to the spectrum, one multiplies the former by the factor $(dW/dt_e)^{-1}$. This is the origin of the spectrum peaks. The spectral divergence at the peaks is evidently integrable since the total number of photons emitted during one period of the revolution is finite. Let us remind also that the obtained spectrum was calculated for a single orbit with a fixed radius. If a radiating domain is a ring of the finite width, one should integrate the spectrum over the radius $\rho_e$ with a weight proportional to the density of the matter of Iron ions in such a ring. After this the infinite peaks disappear and the spectrum would be regularized.

The quantity
\be\n{width}
\Delta=2\Omega Z_m
\ee
determines the width of the spectrum, while the parameter $\gamma^{-1}$ controls the general redshifts of the spectra.

After these general remarks we return to the discussion of samples of the spectra. For illustration we choose the inclination angle $\theta_{o}=25^{o}$. Figure~\ref{F2} shows the spectrum for the ISCO orbit $\rho=6M$ in the absence of the magnetic field.  Figure~\ref{F3} shows a similar spectrum for the same radius of the orbit but when the magnetic field is $b\approx2.251$, while Figure~\ref{F4} shows the spectrum for the ISCO orbit with the same value of the magnetic field $b\approx2.251$.

By comparing Figures~\ref{F2} and \ref{F3} one can see that if one increases the magnetic field keeping the other parameters ($\rho_e$ and $\theta_o$) fixed, then the spectral profiles are narrowed. By comparing Figures~\ref{F2} and \ref{F4} one concludes that for ISCO orbits this narrowing is accompanied by a general redshift of the spectral function. 

To summarize, the common features of the spectrum plots are: (1) the existence of the two sharp peaks at the frequencies $W_{\pm}$; (2) the existence of two branches of the spectrum;  (3) the increase of the average redshift of the spectral frequencies for ISCO with the increase of the magnetic field; (4) the narrowing of the frequency bands with the increase of the magnetic field; (5) the asymmetry of the spectrum with respect to spectral average frequency.
 
The above discussion gives simple qualitative explanations of the properties (1)-(3).
Let us briefly discuss the last two properties. The left plot in Figure~\ref{gamma} shows $\gamma^{-1}$ as a function of the ISCO radius in the presence of the magnetic field $b$. The larger value of the magnetic field, the closer to the horizon is the corresponding ISCO and the greater is the redshift. Numerical calculations confirm also that the width \eq{width} decreases with the increase of $b$ (property (4)). The asymmetry of the spectrum is a generic property of the broadening of the sharp spectral lines for the emitters moving near black holes. It is a result of the relativistic (Doppler) beaming effect. The calculations show that the asymmetry effect becomes more profound when the inclination angle becomes larger.

\section{Discussion}

In this paper we study the effect of broadening of sharp spectral lines emitted by charged particles moving in the vicinity of magnetized black holes. In order to make calculations more transparent we made a number of simplifications. First of all, we restricted ourselves to the case of non-rotating black holes. Certainly, the effects connected with rotation are important. One can expect that astrophysical black holes are rotating, and at least some of them are rapidly rotating with the rotation parameter $a=J/M$ close to 1. For such black holes, even in the absence of the magnetic field, the ISCO radius is located close to the event horizon. In this sense, the action of the rotation on the motion of neutral particles is similar to the effect of the magnetic field on the charged particles motion. In the post-Newtonian approximation the gravitomagnetic field is proportional to the dragging frequency of the local inertial frames. As a result, the gravitomagnetic effects are locally equivalent to inertial effects, that is the gravitational analogue of Larmor's theorem  (see, e.g., \cite{Mashhoon}). However, there exist an important difference. The local velocity at ISCO of a neutral particle in the Kerr geometry tends to the speed of light in the limit $a\to 1$, while its angular velocity $\Omega$ (as measured by a distant observer) remains finite and in this limit it coincides with the angular velocity of the black hole $\Omega_H$. In the non-rotating magnetized black hole the charged particle velocity at ISCO practically does not depend on its position (and hence, on the magnetic field) and is about a half of the speed of light (see the right plot in Figure~\ref{gamma}). As a result, in a strong magnetic field the ISCO angular velocity vanishes. We remind that namely this parameter determines the width of the spectral profiles in the magnetized black holes. This makes it interesting calculations of the broadening spectra in rotating magnetized black holes.

Another simplifying assumption is the form of the magnetic field. In realistic black holes one cannot expect that the magnetic field is homogeneous and extends to infinity. However, for the motion of a charged particle in the equatorial plane and in the black hole vicinity this approximation might be reasonable. It is easy to extend the results for other types of a regular magnetic field, e.g., for the dipolar magnetic field around a static black hole (see, e.g., \cite{PrVa}). Moreover, a model of the homogeneous magnetic field is a good approximation for more realistic magnetic fields generated by currents in a conducting accretion disk, provided the size of the black hole is much smaller than the size of the disk (see, e.g., discussion in \cite{Pet:74}).

In our model we assumed also that charged particles (Iron ions) which emit sharp lines after passing through the radius $6M$ do not fall into the black hole directly, but being supported by the repulsive magnetic field continue their motion in circular orbits. We assume that the radius of these orbits changes slowly (for example, as a result of the synchrotron radiation). So that such charged particles effectively
are accumulated inside $6M$ and are spread up to their ISCO radius. In order to find effective spectrum broadening one should average the obtained expression \eq{SPEC} over such a ring of the finite size. Such averaging would introduce a natural cut-off of the infinite peaks. It would be interesting to estimate the contribution of such ring's radiation to the general broadening of the integrated disk radiation. In order to perform such calculations one needs rather detailed model of the accreting matter structure inside $6M$.

However, one can expect that the main features of this inner domain radiation, namely bigger redshift and narrowing of the spectrum, are of a general nature. This allows one to hope that observations of the broadening of Fe K$\alpha$ lines in magnetized black holes can provide one with the direct information about the magnetic field in the black hole vicinity.

\acknowledgments

We thank Ted Jacobson for his remarks and suggestions proposed during the Peyresq 18 meeting, that stimulated our work on the problem of the spectral line broadening in magnetized black holes. The authors are grateful to the Natural Sciences and Engineering Research Council of Canada for its support. One of the authors (V.F.) thanks the Killam Trust for its financial support.

\end{document}